\UseRawInputEncoding 
\documentclass[final,3p,times,comma,round,numbers,sort&compress,twocolumn]{elsarticle}
\pdfoutput=1

\usepackage{graphicx}
\usepackage[numbers]{natbib}
\usepackage{hyperref}
\usepackage{subfiles}
\usepackage{lineno}
\usepackage{amsmath}
\usepackage[hang,flushmargin]{footmisc} 
\usepackage{xcolor}
\usepackage{caption}
\usepackage[font=small,labelfont=bf,justification=justified,format=plain]{caption}

\renewcommand{\figurename}{Fig.}
\newcommand{\squeezeup}{\vspace{-2.5mm}}
\renewcommand\thesection{\Roman{section}}
\renewcommand\thesubsection{\thesection.\arabic{subsection}}
\usepackage{xpatch}

\xpatchcmd{\MaketitleBox}{\hrule\vskip12pt}{\vspace{-3.5\baselineskip}}{}{}
\xpatchcmd{\MaketitleBox}{\hrule}{\vspace{2\baselineskip}}{}{}

\renewenvironment{abstract}{\global\setbox\absbox=\vbox\bgroup
  \hsize=\textwidth%
  \noindent\unskip\textbf{Abstract}
 \par\medskip\noindent\unskip\ignorespaces}
 {\egroup}
 
\usepackage{multibib}
\usepackage[normalem]{ulem}
\newcites{sec}{references}

\usepackage{etoolbox}
\patchcmd{\abstract}{Abstract}{}{}{}

\journal{ArXiv.org} 

\begin{document}
\renewcommand{\abstractname}{\vspace{-\baselineskip}}
\renewcommand{\theaffn}{\arabic{affn}}

\begin{frontmatter}


\author[add1]{Raymond F. Smith}
\author[add2]{Vinay Rastogi}
\author[add1]{Amy E. Lazicki}
\author[add1]{Martin G. Gorman}
\author[add1]{Richard Briggs}
\author[add1]{Amy L. Coleman}
\author[add1]{Carol Davis}
\author[add1]{Saransh Singh}
\author[add2]{David McGonegle}
\author[add1]{Samantha M. Clarke}
\author[add1]{Travis Volz}
\author[add1]{Trevor Hutchinson}
\author[add1]{Christopher McGuire}
\author[add1]{Dayne E. Fratanduono}
\author[add1]{Damian C. Swift}
\author[add1]{Eric Folsom}
\author[add3]{Cynthia A. Bolme}
\author[add4]{Arianna E. Gleason}
\author[add1]{Federica Coppari}
\author[add5]{Hae Ja Lee}
\author[add5]{Bob Nagler}
\author[add5]{Eric Cunningham}
\author[add5]{Eduardo Granados}
\author[add5]{Phil Heimann}
\author[add1]{Richard G. Kraus}
\author[add1]{Robert E. Rudd}
\author[add6]{Thomas S. Duffy}
\author[add1]{Jon H. Eggert}
\author[add2]{June K. Wicks}

\address[add1]{Lawrence Livermore National Laboratory, P. O. Box 808, Livermore, CA 94550, USA}
\address[add2]{Department of Earth and Planetary Sciences, Johns Hopkins University, Baltimore, MD 21218, USA}
\address[add3]{Atomic Weapons Establishment,
Aldermaston, Reading, RG7 4PR, United Kingdom}
\address[add4]{Los Alamos National Laboratory, Los Alamos, New Mexico, USA}
\address[add5]{Linac Coherent Light Source, SLAC National Accelerator Laboratory, Menlo Park, CA, USA}
\address[add6]{Dept. of Geosciences, Princeton University, Princeton, NJ 08540, USA}


\title{Development of slurry targets for high repetition-rate XFEL experiments\vspace{-2ex}\vspace{1em}}



\address{}

\date{\today}
\begin{abstract}
Combining an x-ray free electron laser (XFEL) with high power laser drivers enables the study of phase transitions, equation-of-state, grain growth, strength, and transformation pathways as a function of pressure to 100’s GPa along different thermodynamic compression paths.~Future high-repetition rate laser operation will enable data to be accumulated at $>$1 Hz which poses a number of experimental challenges including the need to rapidly replenish the target.
~Here, we present a combined shock-compression and X-ray diffraction study on vol\% epoxy(50)-crystalline grains(50) (slurry) targets, which can be fashioned into extruded ribbons for high repetition-rate operation.~For shock-loaded NaCl-slurry samples, we observe pressure, density and temperature states within the embedded NaCl grains consistent with observations for shock-compressed single-crystal NaCl.
\end{abstract}




\end{frontmatter}



\section{Introduction}
\noindent
The use of XFEL sources, which deliver high-flux X-ray pulses (10$^{12}$-10$^{13}$ photons at $\sim$6-25 keV over 50-fs) within a narrow photon-energy bandwidth ($\Delta$E/E $\sim$ 2$\times$10$^{-3}$), has been transformative for the study of dynamic material properties and high energy density (HED) science \citep{Nagler2015}.~Combined with high power laser drivers, this has allowed for measurements previously thought possible only when performed under static compression and integrated in time, including  
complex crystallography \citep{coleman2019,gorman2019,gorman2020,briggs2019b, tracy2019, kim2021}, precise spectroscopy \citep{vinko2015}, and high-resolution imaging \citep{schropp2015,nagler2016,brown2019} in single-shot dynamic compression experiments.

Existing laser drivers at XFELs can shoot every five minutes \citep{brown2017}.~This implies a need for $\sim$100 targets per day, and target fabrication methods have been developed to meet target needs over a five day run. This large data collection capacity has revolutionized the way that high-quality dynamic experiments are done; instead of an experimental campaign spanning months or years at large laser facilities, XFEL HED science campaigns may collect sufficient data in a single day.

The commissioning of the DiPOLE laser at the High Energy Density instrument at the European XFEL (EuXFEL) (100 J in 15 ns, 10 Hz) \citep{mason2018,zastrau2017} will usher in a new era of experiments requiring the delivery of up to 9000 targets over a 15-minute run.~High repetition-rate experiments which combine rapidly adjustable laser pulse shaping with replenishable targets, will allow for high-speed scanning of material properties over large regions of sample pressure-temperature space.
~These capabilities will also allow experimental data to be integrated over long periods of time (many thousands of shots), which will dramatically increase the signal-to-noise over standard single-shot experiments.~This will be of particular importance to low signal phenomena such as liquid diffraction, extended x-ray absorption fine structure (EXAFS) measurements, and scattering off low-Z or low-symmetry materials.

There are three major challenges to overcome for rep-rate experiments: target construction and positioning, data collection and processing, and debris and heat load removal \citep{prencipe2017,orban2020}. To address the requirement for rapid target replenishment it has been proposed to use simple tape targets mounted on a cassette and spooling at 3-5 cm/s \citep{prencipe2017}.~This equates to 30-50 m of tape per 15-minute run. However, the nature of this design requires initial target ductility and so precludes the study of a whole class of brittle ceramics materials.~In addition, the production of extended ribbon targets will produce samples with a characteristic  micro-structural texture and an x-ray texture pattern which may be challenging for Rietveld refinement analysis, and consequently for determination of phase fraction \citep{thompson1987, gorman2020}.

Here we describe the manufacturing and testing of epoxy-sample mixtures or \emph{slurry} targets for use in high-repetition rate experiments on XFEL facilities.~Slurry target designs can produce x-ray diffraction powder patterns for brittle materials, and can be fashioned into extended ribbons for use in high-repetition rate cassette-spool target designs.~A complete description of the sample preparation is given in Appendix A and is briefly described here.~In our slurry targets the material of interest is ground in a mortar and pestle and the resultant powder is sieved through a mesh to ensure individual grain sizes are less than $\sim$1-$\mu$m.~This powder is then thoroughly mixed with Stycast 1266 epoxy in a ratio to obtain a 50\% volume packing and a random distribution of grains.~Large area sheets of slurry material are then formed between polyimide and teflon sheets at the uniform thickness needed for laser-shock experiments ($\sim$30-$\mu$m) by processing through rollers with precision separation.

\subsection{Continuum studies of shock-compressed mixtures}
\medskip
\noindent
There have been extensive studies spanning several decades aimed at determining the continuum response of mixtures \citep{duvall1971, munson1971, krueger1991, kraus2010,  munson1978, bober2019, bober2020, vogler2010, petel2010, montgomery2009, setchell2005} -- and epoxy-sample slurry \citep{munson1978, bober2020, setchell2005, bober2019, vogler2010, petel2010, montgomery2009} in particular -- under an applied shock, in terms of the aggregate stress-strain response  \citep{munson1978, bober2020, setchell2005, petel2010, jordan2012, montgomery2009, bober2019}, the influence of particulate size, packing and distribution \citep{bober2020, montgomery2009}, direct imaging of particulate flow \citep{bober2019}, and the partitioning of energy states within the constituent materials \citep{kraus2010}. A brief overview of these studies is presented here.

\begin{figure}[t!]
	\begin{center}
	\includegraphics[width=1\columnwidth]{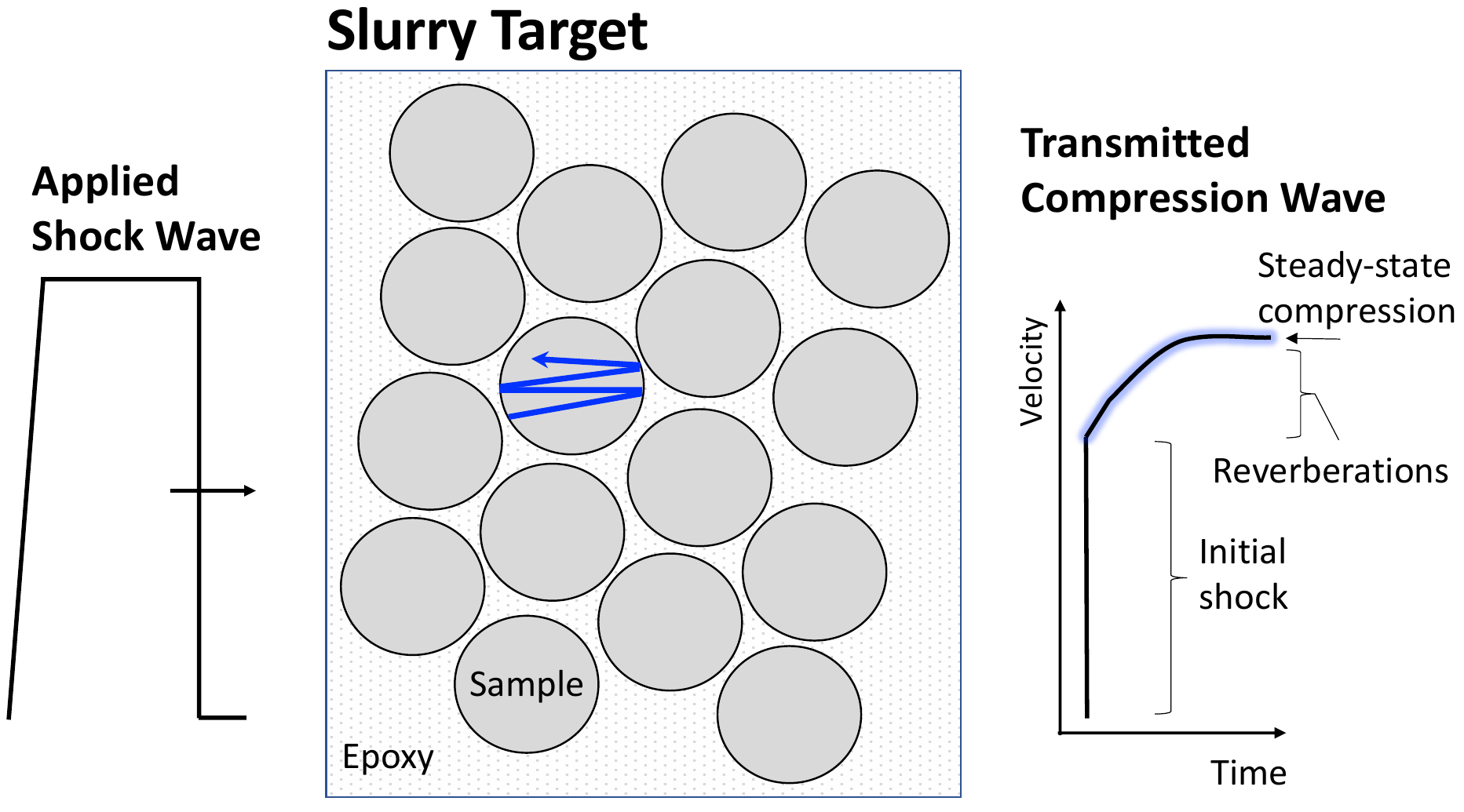}
\caption{Shock-compression of a slurry sample will result in an initial compression along the Hugoniot followed by multiple inter- and intra-grain reverberations towards a steady-state.~Pressure equilibrium between the epoxy and sample is expected to occur rapidly.~Temperature equilibrium, however, is a slower process, achieved through heat flow across sample-epoxy interfaces.
	}
	\label{fig:slurry}
	\end{center}
	\squeezeup
	\squeezeup
\end{figure}

For a random distribution of grains suspended within a cohesive epoxy matrix, and subject to an applied instantaneous shock front, there are a number of physical processes which can modify the thermodynamic compression path \citep{bober2020}.~At low pressures the ``shock" rise time can be broadened by the viscoelastic response of the epoxy \citep{vogler2010}.~If the shock front thickness is comparable to the grain size, the compression of the grain will be isentropic in nature \citep{kraus2010}.~At increasing stress levels, however, the rise-time of the shock wave decreases in a manner consistent with full-density metals \citep{vogler2010, swegle1985} and the shock front thickness becomes small relative to the grain size.~Here, as the shock enters an individual grain there is an instantaneous state reached along the Hugoniot which is dependent on the epoxy-sample impedance difference.~This is followed by multiple inter- and intra-grain wave reverberations until a steady compression state is reached (Fig. \ref{fig:slurry}).~Experimentally this has been observed in transmitted velocity profiles as a sharp rise (due to the shock), followed by a rounded push to a peak pressure state (due to reverberations), over a time period which is dependent on the epoxy-sample impedance-mismatch, the inter-grain compression-wave round trip time, and the granular packing fraction \citep{bober2019, bober2020, vogler2010}.~Wave reverberations are accompanied by the relative motion of the sample grains and the epoxy matrix \citep{bober2019}, where the shear strength of the components can affect the evolving grain distribution (average slurry density).~For high packing fractions inter-grain interactions may also need to be considered in a comprehensive model description of the states generated.~At cessation of the applied shock, the resulting pressure release wave has been observed to be unexpectedly fast compared to the release wave velocity from full density samples \citep{vogler2010}.

While the timescale for epoxy-sample pressure equilibrium is relatively fast and can be constrained by transmitted wave profile measurements, the timescale for temperature equilibrium is longer and difficult to constrain.~Indeed, partitioning of energy between the individual phases is not well understood \citep{kraus2010}. The temperature states reached (epoxy vs sample) are initially dependent on the individual equations-of-state (EOS) but at later times are also dependent on the thermal conductivities of the constituent materials -- i.e. the heat flow between the sample and epoxy.
~In full density samples, inertial confinement conditions and material strength may combine to increase sample temperature due to work heating \citep{Bradley2009}.~In slurry samples, lateral expansion of grains into the compressible epoxy is possible which may reduce the levels of work heating.~We note that diamond-slurry samples were recently used in X-ray diffraction experiments on the National Ignition Facility for laser ramp-compression to 2 TPa, with the goal at reducing sample temperatures at these extreme levels of compression \citep{Lazicki2021}.  


\bigskip
\noindent
For many material-epoxy composites the material states generated by an applied shock has been determined through shock-velocity ($U_s$) -- particle velocity ($u$)  measurements \citep{vogler2010, montgomery2009, bober2019}, and models have been developed which give good qualitative agreement to measured data \citep{vogler2010, jordan2012, petel2010, montgomery2009}. However, these data and models provide information only on the aggregate response.~To qualify the use of slurry targets for future high-repetition rate XFEL target designs it is important obtain an understanding on the compression state within the embedded granular material in terms of density ($\rho$) and temperature ($T$) as a function of pressure ($P$). 

Here, we describe combined laser-shock and X-ray diffraction (XRD) measurements on the Matter in Extreme Conditions (MEC) endstation at the Stanford LCLS XFEL, which provide a direct measure of granular density and crystal structure as a function of increasing shock pressure \citep{gorman2019,gorman2020}.
~Temperature in the sample is constrained by comparing crystal structure evolution as a function of sample density with recent similar data from full-density samples \citep{Rastogi2021}.~Within the accuracy of our data, we find good agreement with the $P$-$T$-$\rho$ states generated within NaCl-slurry samples, with those from single crystal NaCl [100] samples.



\section{Experiments}
\noindent
In these experiments we combine laser-shock compression of a vol\% epoxy(50):NaCl(50) sample with in situ XRD to measure the crystal structure evolution as a function of compression (measured density).

NaCl is an ionic solid of significant interest for high-pressure science, geoscience, and dynamic compression.~It is widely used as a pressure standard and insulating material in diamond anvil cell experiments and there has been much effort devoted to studying it's strength \citep{meade1988, Zhongying2012, mi2018}, equation of state \citep{Fei2007b,Decker1971, Birch1978, Brown1999, Florez2002} and phase transitions \citep{Fritz1971, Nishiyama2003, Li1987, Sata2002, boehler1977, Akella1969, Li2015,Marsh1980}.~At room temperature NaCl undergoes a transition from the B1 rocksalt structure (space group = Fm$\bar{3}$m) to the B2 cesium chloride structure (space group = Pm$\bar{3}$m) at about 25 GPa. The B1-B2 transition is of fundamental interest in Earth science as it is exhibited by oxides such as CaO and MgO at much higher pressure  \citep{Coppari2013, Jung2005}.
~In NaCl, the B1-B2 transition has been extensively studied by static \citep{Nishiyama2003, Sata2002, Li1987, boehler1977, Akella1969, Li2015} and dynamic techniques \citep{Fritz1971, Marsh1980}.~Recent combined XRD-laser shock studies at MEC-LCLS on NaCl [100] single crystal samples \citep{Rastogi2021} have observed transformation into the B2 phase  at 28(2) GPa, B2-liquid coexistence between 54(4) – 66(6) GPa, with near full melt at 66(6) GPa. The states generated over nanosecond laser compression timescales where found to be in agreement with measurements under static compression at $\sim$ten orders of magnitude lower compression rate, consistent with equilibrium conditions being achieved under laser shock conditions \citep{Rastogi2021}.


\smallskip
\subsection{Experimental Setup}
\begin{figure}[h!]
	\begin{center}
	\includegraphics[width=1\columnwidth]{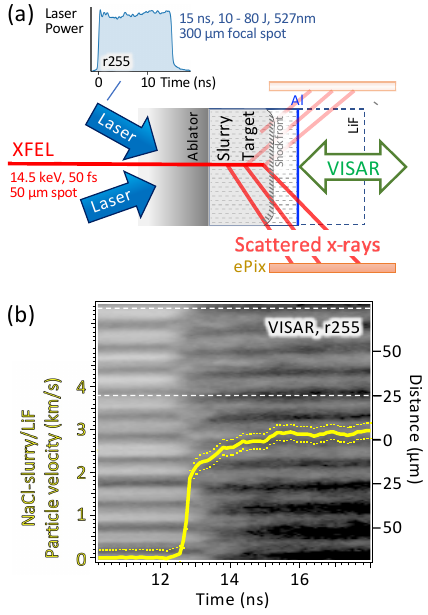}
\caption{\textbf{(a)} Target design for combined laser-shock and x-ray diffraction experiments on LCLS. During shock transit within the slurry layer, the XFEL beam scatters off the compressed NaCl grains and the resultant Debye-Scherrer cones are recorded in transmission geometry on ePix detectors. Also shown is the 15-ns laser pulse shape for MEC shot\# r255. \textbf{(b)} VISAR interferogram for MEC shot\# r255 records target velocity information along a spatially-resolved 1ine at the target plane \citep{Celliers2004}. Here fringe movement is directly proportional to the NaCl-slurry/LiF particle velocity (overlaid yellow curve with standard deviation error bars), integrated over the 50 $\mu$m region of interest (matched to the XFEL beam diameter), shown between the two dashed white horizontal lines. 
	}
	\label{fig:Target}
	\end{center}
	\squeezeup
	\squeezeup
	\squeezeup
\end{figure}

\smallskip
\noindent
The target assembly, as illustrated in Fig. \ref{fig:Target}a, consists of a 50 $\mu$m polyimide [C$_{22}$H$_{10}$N$_{2}$O$_{5}$] ablator layer adhered directly onto a $\sim$30 $\mu$m-thick NaCl-slurry target and, for some shots, a LiF window for velocimetry measurements (Fig. \ref{fig:Target}b).~A 0.1-$\mu$m coating of Al was applied to the LiF to enhance target reflectivity.

\begin{figure*}[!t]
\begin{center}
\includegraphics[width=0.95\textwidth,height=8.5cm]{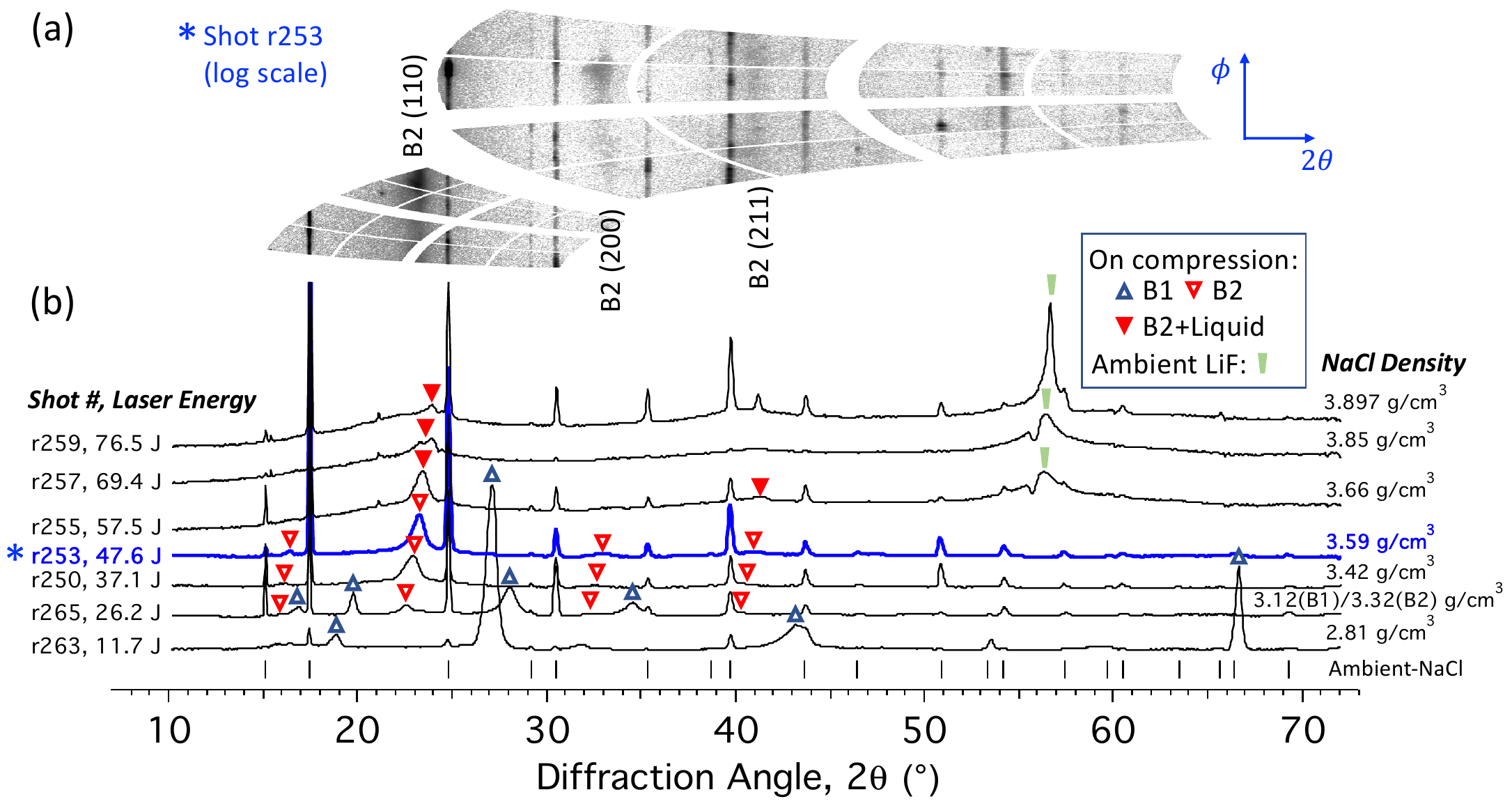}
\caption{\label{fig:Slurry_2021} \textbf{(a)} Raw diffraction data recorded on a subset of the ePix detectors for shot\# r253 and projected into linear 2$\theta$-$\phi$ space. The calibrated 2$\theta$ scale is shown on the bottom axis along with tick marks which denote the peak positions of the ambient pressure B1 phase.~The $\phi$ coverage for the images shown is 80$^{\circ}$. \textbf{(b)} XRD images are integrated -- over all ePix detectors -- to produce the diffraction profiles. Here we show data from seven shots with increasing compression from bottom to top.~As the scattering efficiency for epoxy is low the signal is dominated by diffraction from the NaCl grains.~The integrated profile for the data shown in (a) is plotted as the blue curve.~Each profile contains information on the laser energy (left) and NaCl density (right) based on the measured compressed-sample line positions. We observe a clear evolution with compression from: (1) compressed B1 (shot\# r263), (2) mixed B1+B2 (shot\# r265), (3) mixed B2+liquid (shot\# r255), (4) and a reduction of B2 scattering intensity with compression further into the B2-liquid mixed phase. The evolution of structure with NaCl density is consistent with measurements from shock-compression of full-density NaCl samples \citep{Rastogi2021} (See Fig. \ref{fig:NaCl}). 
}
\end{center}
\squeezeup
\squeezeup
\squeezeup
\end{figure*}

The front surface of the polyimide was positioned within the focal plane of four laser beams which delivered a combined energy of up to 80 J at 527-nm in a 15-ns flat-top pulse within a $\sim$300 $\mu$m diameter focal spot.~Laser energy absorbed by the polyimide ablator causes it to ablate and expand rapidly.~The momentum transfer from this process causes a shock wave, with an initial 15-ns pulse duration, to propagate through the target assembly.~The pressure in the sample and the temporal steadiness of the compression wave are controlled by varying the total laser power and laser pulse shaping, respectively \citep{brown2017}.~An example laser pulse shape is shown in Fig. \ref{fig:Target}a.~A line-imaging velocity interferometer (VISAR) was used to accurately determine the shock arrival time at the slurry free-surface or slurry/LiF interface velocity \citep{Celliers2004}.~In the experiments where a LiF window was used, the VISAR recorded the slurry/LiF particle velocity history ($u$), which was used to constrain the sample pressure during the x-ray probe period. 

The 50-fs output of the LCLS-XFEL at 14.5-keV was incident onto the target during shock transit within the slurry layer, at normal incidence, in a 50-$\mu$m spot centered on the laser drive.~The X-ray beam pointing and timing, relative to the laser-shock drive, is known to a few $\mu$m's and $<$ 100-ps by measuring localized changes in VISAR reflectivity and fringe position, due to the transient change in refractive index of the LiF window caused by charge carrier generation by the x-ray pulse \citep{Rastogi2021}. 
X-rays scattered from the compressed slurry sample were recorded in transmission geometry on several ePix detectors \citep{Dragone2014}.~The angular position of the detectors in 2$\theta$--$\phi$ space was accurately determined by diffraction patterns from ambient pressure Lab6 and CeO$_2$ standards, where 2$\theta$ is the diffraction angle and $\phi$ is the azimuthal angle around the incident x-ray beam \citep{tracy2019}.

\section{Results}
\noindent
Raw X-ray diffraction data obtained under shock compression, and projected into linear 2$\theta$--$\phi$ angular space, is shown, for a subset of the ePix detectors, in Fig. \ref{fig:Slurry_2021}a.~The location of diffraction peaks in 2$\theta$ permits determination of atomic lattice d-spacing, through Bragg's law, and the calculation of sample density.~X-rays are timed to probe the sample during shock transit within the NaCl-slurry layer, and therefore, the recorded diffraction pattern, which is volume-integrated, represents contributions from the shocked, and unshocked regions of the slurry sample.
~Ahead of the shock front, diffraction from the uncompressed NaCl grains ($\rho_0$ = 2.16 g/cm$^3$) -- embedded within the epoxy ($\rho_0$ = 1.12 g/cm$^3$) -- produces a number of sharp ambient pressure peaks associated with the B1 structure (Fig. \ref{fig:Slurry_2021}a). The extent of intensity uniformity in $\phi$ is function of the number of randomly orientated NaCl grains within the volume defined by the 50 $\mu$m diameter X-ray beam and the thickness of the uncompressed slurry sample. Azimuthal intensity variations are due to a non-ideal distribution of grains within the X-ray probe volume.~This has also been observed in our XRD measurements on shock-compressed Al$_2$O$_3$- and MgSiO$_3$-slurry samples (see Supplementary Materials Fig. \ref{fig:XRD_v2}).~We note that although our samples were thoroughly mixed, we did not carry out a statistical analysis of the grain distribution.~This can be improved for future experiments.~A larger X-ray spot, with a more even grain distribution, and/or smaller grains, would produce more powder-like peaks.~We also note that while the starting slurry thickness was $\sim$30-$\mu$m, the region of uncompressed slurry contributing to the diffraction peaks is $< $15-$\mu$m-thick. For future high-repetition rate operation where the XRD data is integrated over many shots to enhance signal-to-noise, this azimuthal intensity variation is expected to smooth out to produce a powder pattern suitable for Rietveld refinement analysis \citep{thompson1987, gorman2020}.

~Shown in Fig. \ref{fig:Slurry_2021}b are a series of $\phi$-integrated intensity profiles (over all ePix detectors) for different shots taken as a function of increasing laser energy (increasing shock pressure).~At low levels of compression ($\rho_{NaCl}$ = 2.81 g/cm$^3$) we observe 
broad peaks of the compressed B1 structure.~We observe similar peak broadening in Al$_2$O$_3$ and MgSiO$_3$ slurry samples (Supplementary Materials Fig. \ref{fig:XRD_v2}).~Peak broadening under shock compression is generally attributed to grain-size reduction due to plastic/brittle deformation \citep{milathianaki2013, hawreliak2008}.~An additional contributor to peak broadening is the distribution of pressure states within the sample (see Fig. \ref{fig:Impedance}).~At higher compression (shot\# r265, $\rho_{NaCl}$ $<$ 3.12$\rightarrow$3.32 g/cm$^3$), we observe diffraction peaks consistent with a mixed B1+B2 phase assemblage. 

For $\rho_{NaCl}$ = 3.42$\rightarrow$3.59 g/cm$^3$ only compressed B2 is observed.~While not shown in Fig. \ref{fig:Slurry_2021}, we note that at late times, and along a pressure release path from an initially compressed state within the B2 phase, we observe the reverse B2$\rightarrow$B1 transition.

~For $\rho_{NaCl-B2}$ = 3.59$\rightarrow$3.90 g/cm$^3$ a mixed B2+liquid assemblage is observed.~Increased compression over this density range results in a diminished intensity of the B2 peaks, and an increased broad background intensity from liquid diffraction.~This is consistent with the expected phase evolution as the Hugoniot crosses the melt line \citep{gorman2015, Rastogi2021}.

\section{Discussion}
\subsection{Pressure Determination}
\smallskip
\noindent
Constraints of sample pressure in laser-shock experiments typically rely on measurements of the sample/LiF interface velocity, and standard impedance-matching techniques.~This approach requires an knowledge of the $P$-$u$ Hugoniot relations for both the sample and LIF window. While there have been extensive experimental and theoretical studies aimed at determining the continuum response of exoxy-sample mixtures under an applied shock load \citep{vogler2010, montgomery2009, bober2019, jordan2012, petel2010}, the determined Hugoniots represent the average bulk response and therefore do not give direct information on local granular density states. However, due to the multiple wave reverberations in the slurry sample (Fig. \ref{fig:slurry}), sample-epoxy pressure equilibrium is expected to be rapidly obtained, and impedance-matching can then be used to constrain the sample pressure.

A review of different theoretical models for determining the composite Hugoniot response for mixtures is given in Ref. \citep{petel2010}. The models all assume that the internal energy and density of the mixture must be related to the weighted sum of the individual component properties. 
Here we consider the simplest of those models - which assumes that the mixture is under pressure equilibrium and that an equilibrium material velocity of the mixture can be determined by averaging the material velocities on the individual component Hugoniots according to a velocity based mixture rule \citep{petel2010}. Applying this approach the vol\% epoxy(50):NaCl(50) slurry used in our experiments, the mixture velocity is obtained for a given pressure from the following relation,

\begin{equation}
u_{slurry}^2 = X_{NaCl}\:  u_{NaCl}^{2}+(1-X_{NaCl})\: u_{Epoxy}^{2}        ,
\label{eq:Mixed}
\end{equation}
\noindent
where $u$$_{NaCl}$ and $u$$_{Epoxy}$ are the particle velocities for the NaCl and epoxy components, respectively, at a given pressure on their respective individual Hugoniots.~$X$ represents the mass fraction of individual components.~For the vol\% epoxy(50):NaCl(50) slurry, $X_{NaCl}$ = 0.66. This expression is shown to give a good representation of slurry target Hugoniot relations in many composite materials materials (see Supplementary Materials Fig. \ref{fig:Hugoniots} and Ref. \citep{petel2010}).

\begin{figure}[!t]
\begin{center}
\includegraphics[width=1\columnwidth]{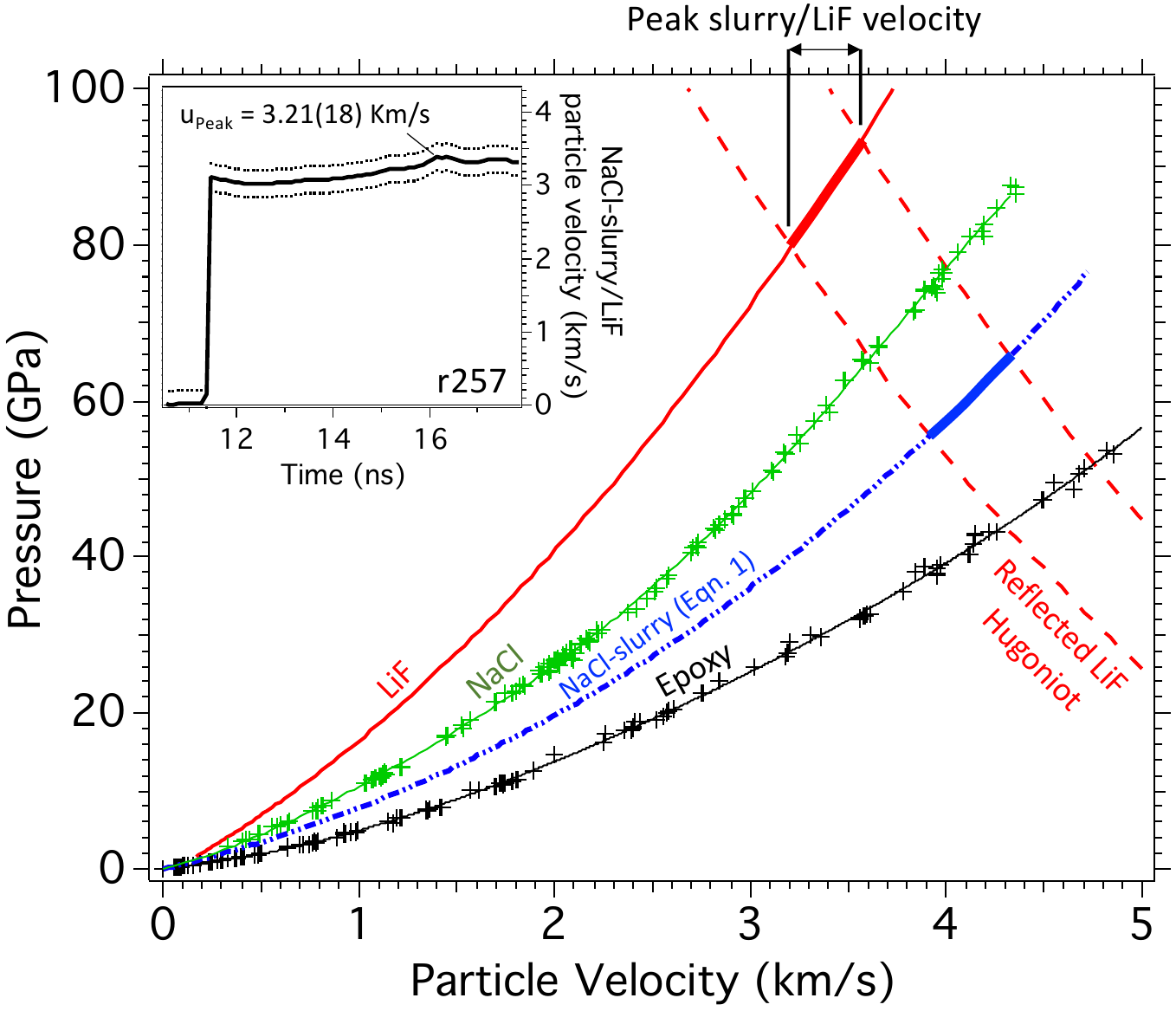}
\caption{Hugoniot $P$-$u$ relations for LiF (red curve: Sesame EOS table\# 7271v3) \citep{davis2016}, NaCl (green curve: data + fit) \citep{Marsh1980, Fritz1971}, vol\% epoxy(50):NaCl(50) slurry (blue curve), and epoxy (black curve: data + fit) \citep{mcqueen1970b, munson1972}. The NaCl-slurry $P$-$u$ curve is derived from Eqn. \ref{eq:Mixed}. The inset figure shows the determined NaCl-slurry/LiF particle velocity for shot\# r257 with error bars which represent the distribution of velocity states over the VISAR spatial region-of-interest (e.g., Fig. \ref{fig:Target}b). The range of measured velocities is shown on the top axis and is highlighted by the bold red curve on the LiF $P$-$u$ Hugoniot.~Estimates of pressure in the NaCl-slurry (bold blue curve) are obtained by the intersection of the reflected LiF Hugoniot (dashed red).~For shot\# r257 the estimated pressure of the NaCl grains is 60.7(7.2) GPa. This value is plotted against the XRD-measured density in Fig. \ref{fig:NaCl} -- in good agreement with previous XRD measurements on single crystal NaCl samples.
}
\label{fig:Impedance}
\end{center}
\squeezeup
\squeezeup
\end{figure}

In Fig. \ref{fig:Impedance} the $P$-$u$ curves for LiF, NaCl and epoxy are shown along with the calculated composite response of the vol\% epoxy(50):NaCl(50) slurry based on Eqn. \ref{eq:Mixed}. Using the measured slurry-LiF particle velocity (e.g. Fig. \ref{fig:Target}b and inset to Fig. \ref{fig:Impedance}), and standard impedance matching, the pressure in the NaCl grains may be determined.

\begin{figure}[!t]
\begin{center}
\includegraphics[width=1\columnwidth]{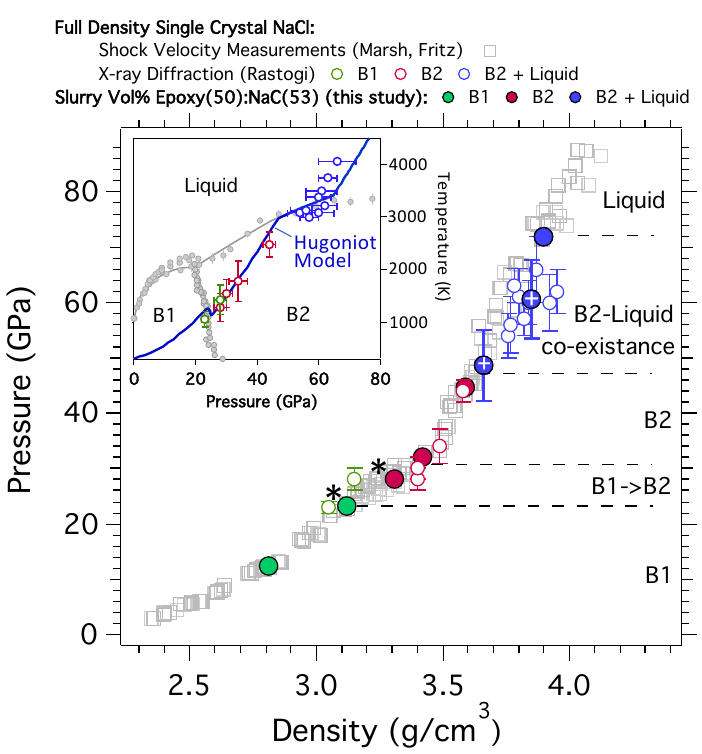}
\caption{
Hugoniot pressure versus density measurements on NaCl: $U_s$-$u$ measurements on single-crystal samples with no structural determination (open squares \citep{Marsh1980,Fritz1971}), XRD measurements on single-crystal samples (open circles color-coded to denote measured crystal structure \citep{Rastogi2021}).~The densities determined from shock-loaded NaCl-slurry targets, and fixed to the known Hugoniot pressure \citep{Marsh1980, Fritz1971}, are shown as filled circles with color representing the measured high-pressure phase (Fig. \ref{fig:Slurry_2021}).~Single shot data showing a mixed B1+B2 phase assemblage is denoted by *.~Two NaCl-slurry shots (blue circles with white crosses) employed LiF windows and the pressure + uncertainty values are based on the analysis described in Fig. \ref{fig:Impedance}.~The inset figure shows the $P$-$T$ phase map of NaCl with B1, B2 and liquid stability regions defined by quasi-static compression measurements (grey circles \citep{boehler1977, Nishiyama2003, Akella1969, Li1987}).~The calculated Hugoniot path is shown by the blue curve along with temperature estimates from a recent LCLS-MEC study on shock-compressed single crystal NaCl \citep{Rastogi2021}.~The onset of melt and total melt are expected along the Hugoniot at pressures and densities of 47-65 GPa and 3.62-3.76 g/cm$^3$, respectively. We observe good agreement between NaCl-slurry and single crystal datasets in terms of crystal structure evolution with NaCl density which is consistent with comparable $P$-$T$-$\rho$ compression paths. 
}
\label{fig:NaCl}
\end{center}
\squeezeup
\end{figure}

In Fig. \ref{fig:NaCl} previously reported pressure-density Hugoniot values for single crystal NaCl (open squares, colored open circles) are plotted alongside values obtained from the current NaCl-slurry study (colored filed circles).~For the majority of the NaCl-slurry shots no LiF window was used, and therefore a pressure determination based on impedance matching (as in Fig. \ref{fig:Impedance}) was not possible. For these shots the measured density values were fixed to the pressures determined from the full-density NaCl Hugoniot measurements \citep{Marsh1980}.~For two slurry shots LiF windows were used and pressure was determined though impedance matching (white crossed symbols in Fig. \ref{fig:NaCl}). These pressure-density points are in in good agreement with previous XRD measurements on single crystal NaCl samples \citep{Rastogi2021}.

We note that in the model depicted in Fig. \ref{fig:slurry}, the final steady-state crystalline-sample density -- along the combined shock + reverberation path -- is expected to be higher than the Hugoniot at an equivalent pressure. However, within the accuracy of our measurements this is effect is not apparent.

\subsection{Temperature Determination}
\smallskip
\noindent
While we don't get any direct temperature measurement from our data, we can compare with previous data on NaCl single crystals \citep{Rastogi2021} to constrain the temperature evolution of the NaCl slurry grains as a function of pressure.~The inset to Fig. \ref{fig:NaCl} shows the $P$-$T$ diagram for NaCl with stability regions for B1, B2 and liquid, as defined by static compression data (gray circles \citep{boehler1977, Nishiyama2003, Akella1969, Li1987}).~Also plotted is the model Hugoniot and estimated shock temperatures from a recent shock compression study from single crystal NaCl samples (colored open circles \citep{Rastogi2021}). Along the Hugoniot NaCl transforms from B1$\rightarrow$B2 at $T$ $\sim$1200K and melt initiates at $T$ $\sim$3000K. The consistency of phase evolution with compression between the NaCl-slurry data and NaCl single crystal data (Fig. \ref{fig:NaCl}) indicates a comparable thermodynamic compression path.



    \begin{figure}[t!]
	\begin{center}
	\includegraphics[width=0.8\columnwidth]{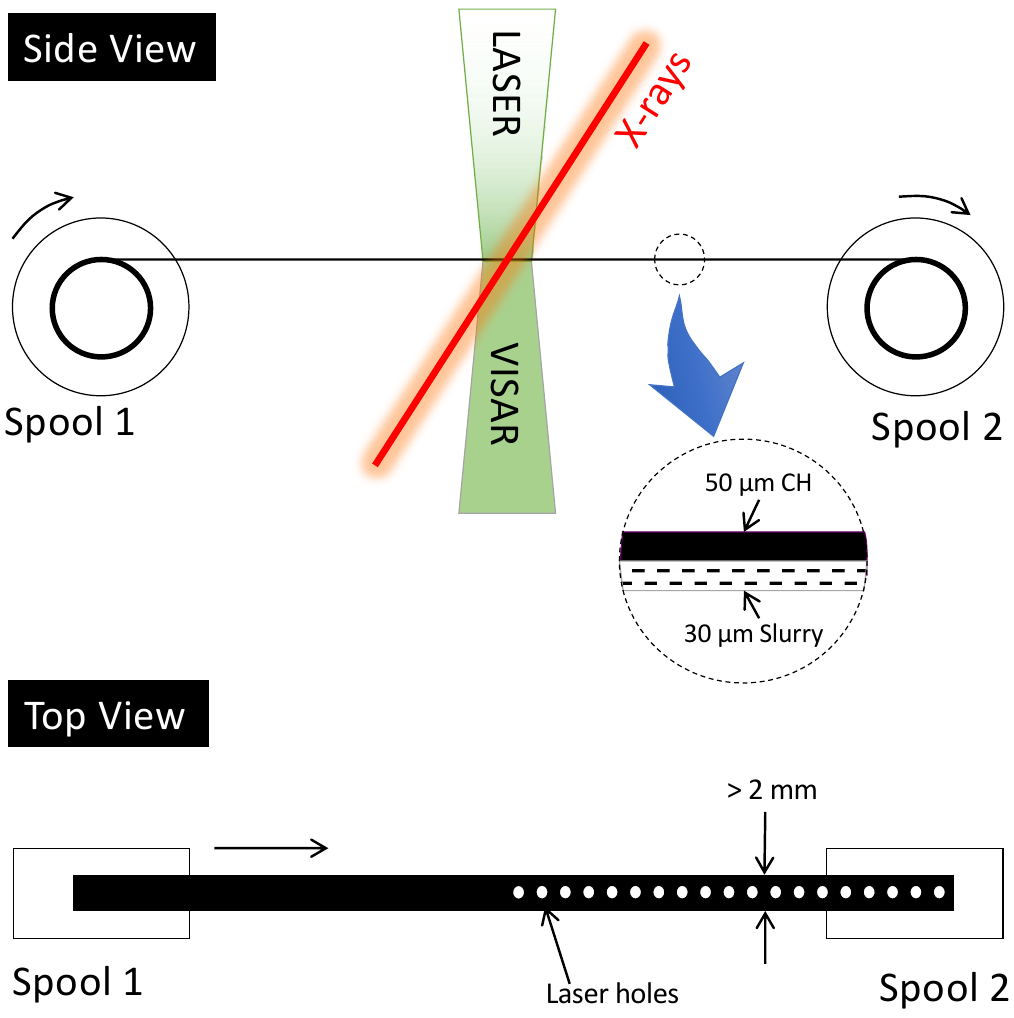}
	\caption{Conceptual design for high repetition-rate replenishable targets which facilitates both laser compression and x-ray diffraction.~A target ribbon of the slurry samples described here may be circulated between two spools within the focal plan of the laser drive beam.~Integration over many shots will increase signal levels and average out any azimuthal intensity variations within the XRD pattern (Fig. \ref{fig:Slurry_2021} and Supplementary Materials Fig. \ref{fig:XRD_v2}).}
	\label{fig:RepRate}
	\end{center}
	\squeezeup
\end{figure}

\section{Conclusions}
\noindent
We have demonstrated the ability to produce powder diffraction patterns from crystalline brittle materials by suspending the sample within a ductile epoxy matrix. This opens up the possibility to study brittle samples in high-repetition rate laser operation on XFEL facilities using a cassette design as illustrated in Fig. \ref{fig:RepRate}. In addition, the transition to slurry targets will give closer analogues to real-world samples like meteorites or rocks that are porous and multi-component in nature.

High-repetition rate diffraction from slurry samples would produce high-quality powder diffraction which lends it self to Rietveld refinement analysis from which phase fraction information can be extracted \citep{gorman2020}.~Our diffraction data on shock-compressed vol\% epoxy(50):NaCl(50) slurry is consistent with the NaCl grains being compressed close to the full-density Hugoniot.~We observe phase evolution from B1, B2 and liquid with increasing pressure.~Future work is needed to accurately determine the thermodynamic compression path for different composite mixtures, through, for example, Debye-Waller analysis of XRD data \citep{debye1913,murphy2008}.




\medskip

\appendix
\section{Procedure for making slurry samples}
\noindent
For hard spheres there is a maximally random jammed (MRJ) packing fraction of about 0.64 \citep{Torquato2000}.~For slurry samples with this packing fraction the remaining 0.36 of the volume would be filled by epoxy.~However we find that for granular volumes $\geq$ 50$\%$, viscosity increases strongly which makes it difficult to mix.~A 50$\%$ epoxy:50$\%$ sample volume ratio is ideal for maximizing the packing fraction while keeping the viscosity low enough for thorough mixing.~The procedure we employed to produce the slurry samples used in this study is described below.

\begin{figure}[!t]
	\begin{center}
	\includegraphics[width=0.8\columnwidth]{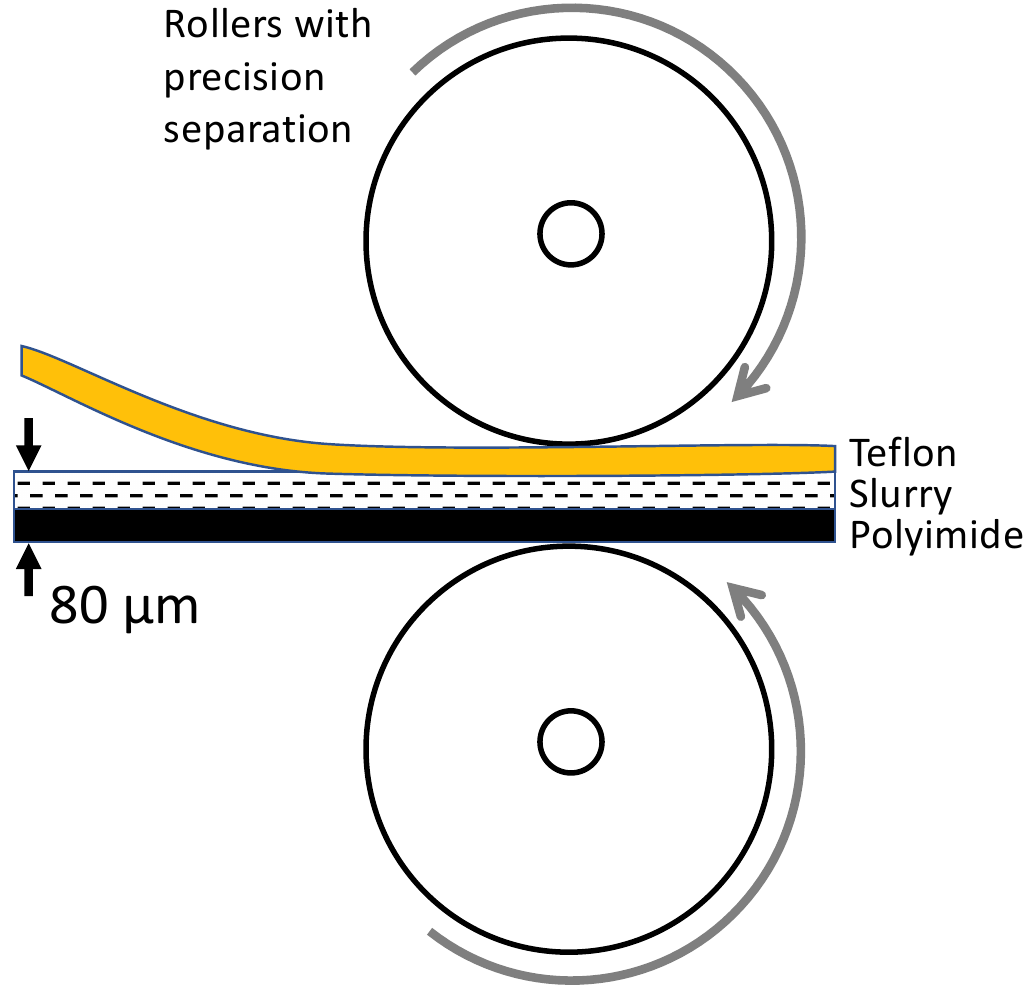}
	\caption{Setup for producing slurry samples at the required thickness for high-repetition rate laser experiments.
	}
	\label{fig:setup}
	\end{center}
	\squeezeup
\end{figure}

\begin{itemize}
\item Crystalline samples are ground to a fine powder, with a mortar and pestle, and sieved through a mesh to ensure individual grains $<$ 1$\mu$m.
\item The powder is then fully baked in a vacuum oven to remove all moisture (otherwise, in the case of diamond powder \citep{Lazicki2021}, the final mixture is visibly heterogeneous).
\item 
Stycast 1266 was chosen for the epoxy ($\rho$$_0$=1.12 g/cm$^3$) due to it's transparency, low viscosity, working time, and shelf life (clear, 650 cP, 30 min., days in a freezer) \citep{barucci1999}.~The viscosity can be further lowered by increasing the temperature, but that shortens the working time.
\item The epoxy must be vacuum de-aired to ensure a void-free embedment, as is necessary for this application.
~This is achieved by placing the epoxy in a bell jar while applying a vacuum.~Initially the glue will foam up as the air bubbles escape.~After a period of time visual inspection will confirm the epoxy is bubble free. 
\item The desired amount of epoxy is placed onto a small tray using a syringe.~With a 0.01-mg accuracy scale the mass is then recorded and the epoxy volume calculated.
\item The commensurate mass of the powdered sample -- to achieve the desired epoxy-sample volume ratio -- is then added to the epoxy. The epoxy+sample components are then thoroughly mixed to ensure a random distribution of grains. 
\item The epoxy mixture may then be fashioned into a planar film as illustrated in Fig. \ref{fig:setup}.~Here the slurry is placed between a polyimide layer and a teflon layer and passed through rollers with a separation to produce the target thicknesses described in Fig. \ref{fig:Target}a.~From the resultant sheet of polyimide + slurry, targets for laser-shock experiments may be punched out ($\sim$2-mm diameters).~Following this approach large areas of slurry samples may be produced as is needed for future high-repetition rate target designs (Fig. \ref{fig:RepRate}).
\end{itemize}

\bibliographystyle{apalike}
\bibliography{main}


\bigskip

\noindent
\footnotesize{\textbf{Acknowledgements:} This work was performed under the auspices of the US Department of Energy by Lawrence Livermore National Laboratory under contract number DE-AC52-07NA27344. This work was supported through the Laboratory Directed Research and Development Program at LLNL (project no.’s 17-ERD-014, 21-ERD-032).~C.A.B. would like to acknowledge support from Science Campaign 2 at Los Alamos National Laboratory, which is operated for the National Nuclear Security Administration of the U.S. Department of Energy under Contract No. DE-AC52-06NA25396.~Use of the Linac Coherent Light Source (LCLS), SLAC National Accelerator Laboratory, was supported by the U.S. Department of Energy, Office of Science, Office of Basic Energy Sciences, under Contract No. DE-AC02-76SF00515.~The MEC instrument is supported by the U.S. Department of Energy, Office of Science, Office of Fusion Energy Sciences, under Contract No. SF00515. The code used for the XRD analysis is publicly available at https://github.com/HEXRD.}


\noindent

\setcounter{equation}{0}
\setcounter{figure}{0}
\setcounter{table}{0}
\makeatletter
\renewcommand{\theequation}{\arabic{equation}}
\renewcommand{\figurename}{Supplementary Figure}
\renewcommand{\thefigure}{\arabic{figure}}
\renewcommand{\bibnumfmt}[1]{[#1]}
\renewcommand{\citenumfont}[1]{#1}
\renewcommand{\thepage}{\arabic{page}} 

\captionsetup[table]{name=Supplementary Table ,labelsep=period}
\captionsetup[table]{labelfont=bf}

\captionsetup[figure]{name=Supplementary Figure ,labelsep=period}
\captionsetup[figure]{labelfont=bf}

\bibliographystylesec{apalike}






\onecolumn
\clearpage

\renewcommand{\thefigure}{S\arabic{figure}}

\renewcommand{\thefigure}{S\arabic{figure}}
\renewcommand{\thetable}{S\arabic{table}}
\renewcommand\thesection{S\arabic{section}}
\renewcommand\thesubsection{S\thesection.\arabic{subsection}}

\captionsetup[table]{name=Table,labelsep=period}
\captionsetup[Table]{labelfont=bf}

\setcounter{table}{0}
\setcounter{figure}{0}
\setcounter{table}{0}
\setcounter{section}{0}
\normalsize 







\noindent
\textbf{SUPPLEMENTARY MATERIALS}

\bigskip
\bigskip
\begin{figure}[h]
\begin{center}
\includegraphics[width=0.95\columnwidth]{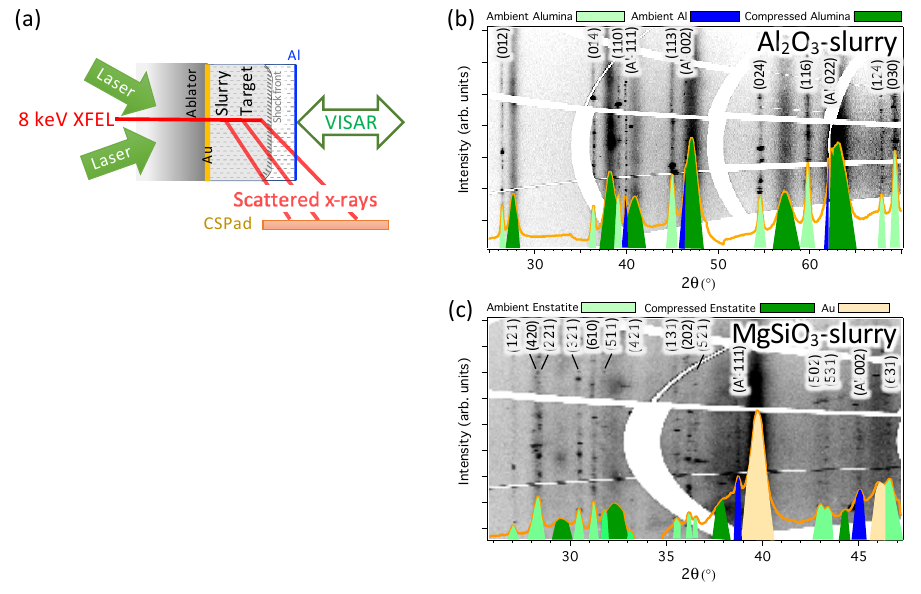}
\caption{Scattered X-ray diffraction during shock compression of \textbf{(b)} alumina (Al$_2$O$_3$), \textbf{(c)} and enstatite (MgSiO$_3$) slurry samples (vol\% epoxy(50):sample(50)) for target design in \textbf{(a)}.~In these experiments no VISAR optical window was used so it was not possible to determine the sample pressure.~For these shots, the 50-fs output of the LCLS-XFEL at 8-keV was incident onto the target, at 24 degrees from target normal, in a 20-$\mu$m spot centered on the laser drive.~The enstatite target had a $\sim$1-$\mu$m Au layer deposited between the polyimide and slurry layer.~The foreground in \textbf{(b)} and \textbf{(c)} shows the azimuthally-integrated profiles with the color-coded assignment of the individual phases.~(\emph{hkl}) assignments for the ambient pressure phases are shown.~In both cases the 50-fs XFEL pulse probes the sample during shock transit, therefore the XRD pattern represents the combined scattering from the compressed and uncompressed slurry volumes.~In \textbf{(b)} the Al$_2$O$_3$ grains ($\rho_0$ = 3.92 g/cm$^3$) within the slurry exhibit a number of sharp ambient-pressure peaks associated with the trigonal phase (space group = R$\bar{3}$c).~Paired with each peak is a broader peak at higher 2$\theta$ angle -- from the shock-compressed slurry volume -- and with a more uniform azimuthal intensity distribution.~Analysis of the XRD data sets the density of the compressed alumina grains at 4.5 g/cm$^3$. In \textbf{(c)} the MgSiO$_3$ grains ($\rho_0$ = 3.2-3.24 g/cm$^3$, assuming ~5mol.\% iron content) within the slurry exhibits a number of sharp ambient-pressure peaks associated with the orthopyroxene structure (space group = Pbca).~Also shown are peaks consistent with the compressed Au layer, ambient-pressure Al, and broad peaks of compressed MgSiO$_3$ with an estimated density of 3.32 g/cm$^3$. We observe an azimuthal intensity variation in the ambient-pressure diffraction peaks.~We adjudge this to be due to the a non-ideal distribution of grains within the slurry volume defined by the $\sim$20-$\mu$m diameter X-ray probe region.}
	\label{fig:XRD_v2}
	\end{center}
	\squeezeup
\end{figure}

\begin{figure*}[h]
\begin{center}
\includegraphics[width=1\textwidth,height=5cm]{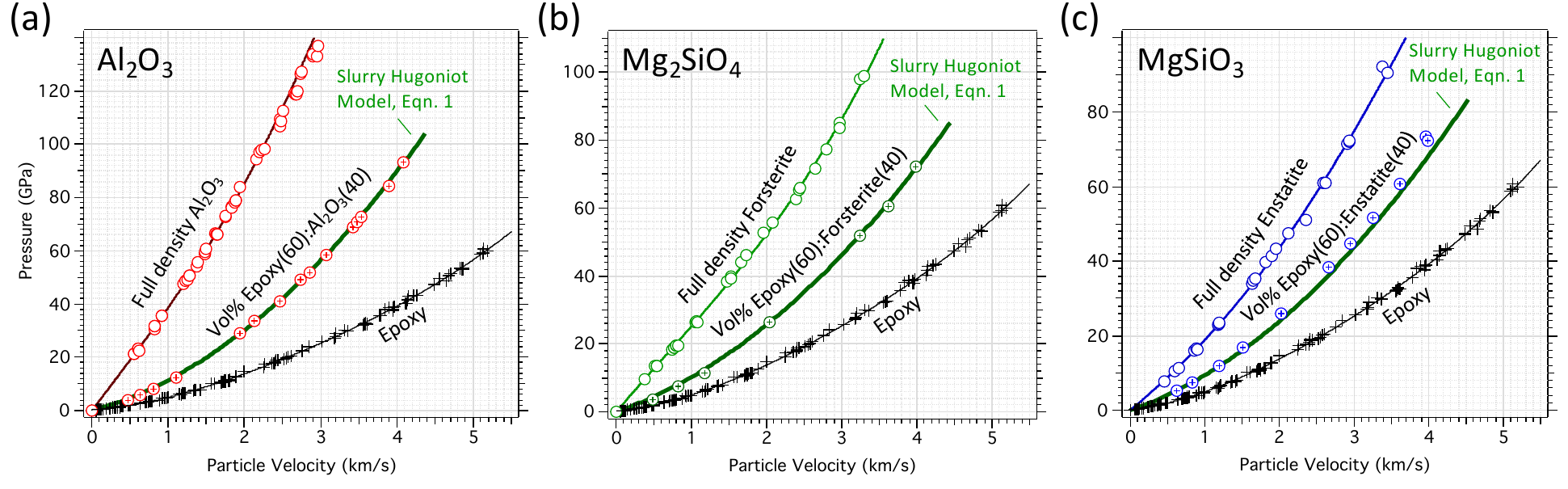}
\caption{Previously published Hugoniot $P$-$u$ data for \textbf{(a)} Al$_2$O$_3$, \textbf{(b)} Mg$_2$SiO$_4$, and \textbf{(c)} MgSiO$_3$ (open circles with fits) \cite{Marsh1980, erskine1994}. Also shown are Hugoniot data for epoxy (black cross with fits) \cite{mcqueen1970b, munson1972}, and slurry sample with vol\% epoxy(60):sample(40) for each material (crossed circles) \cite{Marsh1980}. We observe that, for each material, the calculated slurry Hugoniot based on Eqn. 1 (bold green curve) gives reasonable agreement to the slurry Hugoniot data. The slight disagreement between the model and the MgSiO$_3$ slurry data in (c) may be due to different assumed densities in the model as the density values where not clearly reported in the original study. Good agreement with the slurry Hugoniot model has also been reported with many other composite mixtures \cite{petel2010}.
}
\label{fig:Hugoniots}
\end{center}
\squeezeup
\squeezeup
\squeezeup
\end{figure*}

\end{document}